# Coalition Formation: Concessions, Task Relationships and Complexity Reduction

Samir Aknine
LIP6, Paris 6
8 Rue du Capitaine Scott
75015, France
*samir.aknine@lip6.fr*

Onn Shehory
IBM Haifa Research Lab, Haifa University
Haifa 31905, Israel
*onn@il.ibm.com*

## Abstract

*Solutions to the coalition formation problem commonly assume agent rationality and, correspondingly, utility maximization. This in turn may prevent agents from making compromises. As shown in recent studies, compromise may facilitate coalition formation and increase agent utilities. In this study we leverage on those new results. We devise a novel coalition formation mechanism that enhances compromise. Our mechanism can utilize information on task dependencies to reduce formation complexity. Further, it works well with both cardinal and ordinal task values. Via experiments we show that the use of the suggested compromise-based coalition formation mechanism provides significant savings in the computation and communication complexity of coalition formation. Our results also show that when information on task dependencies is used, the complexity of coalition formation is further reduced. We demonstrate successful use of the mechanism for collaborative information filtering, where agents combine linguistic rules to analyze documents' contents.*

## 1 Introduction

Coalition formation mechanisms are often used as a means for agent coordination. Coalition formation is necessary when agents need to perform tasks which they cannot carry out efficiently alone. Several coalition formation models have been suggested to date (e.g. [5,6,7]), based on various assumptions. Recently, a solution that suggests that agents compromise their gains to promote coalition formation was suggested [3]. In this article, we address the coalition formation problem and suggest compromise too. However, prior work assumes that the value at which compromise is beneficial is known, or can be arrived at experimentally. In many real applications where coalitions are necessary, this assumption does not hold. In our coalition formation mechanism, we do not assume that the optimal compromise point is known in advance. Rather, we provide agents with means to gradually arrive at an agreed compromise via a series of concessions. Such a solution is more flexible and more applicable to real problems compared to previous solutions. Our coalition formation method utilizes dependence relations among tasks to guide agent search for coalitions. Searching the coalition space based on task combinations serves as a means for agents to arrive at compromise. Our approach, in difference from other coalition formation solutions, suggests that agents first attempt to reduce conflicts amongst themselves, and only then form the coalition. Conflict reduction in turn reduces negotiation time between agents, as our experimental evaluation explicitly shows. In addition, our solution considers the complex relationships among tasks in the search for coalitions. Such use of task complexity to improve the coalition formation is not present in other solutions, as task relationships are seldom considered.

Several reasons accredit the use of task structure analysis for coalition search in the multi-agent system. The presence of task relationships, for instance dependencies between tasks, introduces a certain form of dependence between the agents which will perform them. Consequently, dependencies between the coalitions formed by the agents may arise. When these dependencies exist, it would be preferable that agents identify them in advance, as such knowledge can promote compromise and simplify coalition formation. Thus, it should be advantageous to manage dependencies between tasks prior to their assignment to agents. This implies that agents must reason about the tasks they will perform prior to making a decision on the coalitions they join. Dependencies between tasks can be revealed prior to coalition formation via search. Leaving dependencies between tasks unidentified does not avoid conflicts, it merely postpones their resolution. When conflict resolution is postponed to a late stage of the coalition formation process, agents might form coalitions that they will need to break when conflicts are found, thus require further negotiation and re-formation.

Another significant reason for searching the coalition space based on task dependencies is the reciprocal interest of the agents. For instance, an agent *A* may take part in the execution of a task *T*; task *T* may be of significance for another agent *B*; Agent *A* might need *B*'s support in future tasks. If *A* and *B* know this in advance, they are motivated to be more cooperative, and in turn apply compromise regarding the execution of *T* and other tasks.

In this article we illustrate the method we propose in the context of a European project. The project provides a multilingual system for the analysis and detection of racist and revisionist content on the Internet. In this application domain, agents are used to dynamically combine linguistic rules for document analysis. Agents, each implementing a linguistic rule, form coalitions. The coalition formation method consists of clustering tasks into task partitions based on relationships identified among the tasks. Task relationships we consider include dependence, similarity, covering, etc. Task clustering simplifies coalition formation and consequently reduces the negotiation time between agents.

This article is organized as follows. Section 2 introduces the coalition formation problem and the document filtering application context for which we solve it. In section 3, we present the concepts of the coalition formation model, first the principles of this model, then the behaviors of the agents in the model. Section 4 presents the results of the experiments carried out. We then conclude and discuss future work in section 5.

## 2  The problem

In this section we formally define the coalition formation problem we solve and describe the context of the solution.

### 2.1 Problem description

Consider a set of agents $C = \{C_1, C_2, ... ,C_n\}$ and a set of tasks $T = \{T_1, T_2, ... ,T_m\}$. The agents in $C$ need to execute the tasks in T and can negotiate over their execution. Several relationships are defined on $T$. Tasks in $T$ can be combined into sets of tasks, and negotiation can be performed over such sets. Each agent in $C$ has its own utility function that it tries to maximize. The goal of an agent is to determine the coalitions to which it should belong and the tasks or task sets to perform in these coalitions, such that its utility is maximized.

### 2.2 Application context

The application addressed in this study concerns web document filtering. The tracking of racist documents on the Internet faces a number of obstacles, which make it impossible to rely only on the classical keyword-based approach, neural network techniques, etc. [1,4] Racist discourse spans from hate speech to more subtle insinuations, with different themes: racist, revisionist, anti-semitic, etc. Different genres are used in these documents pseudo-scientific articles, pamphlets, etc.
From the analysis of large sets of racist, anti-racist and non-racist documents, a number of criteria for identifying racist content have been identified by the teams of linguists working on the project: (1) *Unique racial expressions* used only by racist people, for example "Rahowa" standing for "Racial Holy War"; (2) *Average frequencies* of certain words in racist documents differ from their average in general documents. These words are not necessarily racist ones but more common words (like "their" or "white"); (3) *Combined frequencies* of certain word pairs are relevant, e.g. the combination of "our" with "civilization", "race" or "religion"; (4) *Suffixes* like "al", "ence", "ism" are good indicators for separating racist and anti-racist documents.

One difficulty of the document filtering system we try to build is that a single criterion is not sufficient for indicating racism; convergence of several criteria is required for such indication; however, this indication may be valid only provided that there are no concomitant indications of anti-racism. Hence the number of criteria (several thousands), their correlations and relative relevance, increase the overall complexity. To address this problem, we introduce a solution based on the use of a multi-agent system. In particular, we employ a coalition formation model to support criteria combination for the information retrieval and filtering problem at hand.

## 3  Coalitions of criteria agents

As stated above, the analysis of documents in the problem domain addressed here employs multiple evaluation criteria. In practice, each document should be evaluated based on a combination of several linguistic criteria. Since combinations vary across documents, and the right combination for each document is not known in advance, there is a need for a dynamic, flexible mechanism for criteria selection and combination. In this work this need is addressed via multi-agent coordination. We introduce a set of *criteria agents*, each representing a single linguistic criterion, and provide them with a dynamic coordination mechanism – coalition formation – to address the criteria combination problem. Following, we present the coalition formation model used by the criteria agents. Each agent, given a document, has to produce a set of characteristics of that document, using both its own processing methods and information produced by other criteria agents. To this end, the important contribution of our work is in the collaborative use of information produced by multiple agents. This collaboration improves the quality of document and site classification compared to results achieved without such collaboration.

### 3.1 Model preliminaries

Based on a single criterion, an agent cannot individually provide a definitive judgment on a document. Hence, agents must dynamically join together to produce definitive judgments. In our solution, this is achieved by agents forming coalitions to analyze documents. A coalition is a temporary association between agents to reach joint goals. For the application domain addressed here, the purpose of a coalition of agents is to categorize a document as racist, revisionist, anti-racist, etc.

The details of a coalition formation protocol depend on the type of problem studied. For instance, varying trust relations or agents' objectives, might require different protocols. To enable the agents to form coalitions, most current protocols (an exception is found in [2,3]) assume that the utility functions of agents, which measure their degree of satisfaction with each suggested solution, must be comparable or the same. This means that agents must be able to agree (for each task or task combination) on a common utility function. This assumption is acceptable for many multi-agent systems, in particular for economic cases where utilities can often be calculated in terms of profit. However, in many cases comparing the utilities of agents, and even more so their aggregation, is nontrivial. The numerical measurement of the utility of an agent is already a strong assumption for itself.

In this work we propose a new coalition formation model which does not require aggregation of agent preferences and utilities. This model does not force the agents to follow a particular order while participating in the coalition formation process; it guaranties an equitable processing of agents' choices. The protocol suggested is particularly suitable for problems with complex tasks (where there is a need for several agents and for coalitions) and for dynamic scenarios where tasks may be added and others cancelled or modified, and where the agents have different utility functions. Agents are self-interested, i.e. they do not necessarily trust each other. However they respect all the commitments to which they agreed. We assume that the utility functions of the agents are not known by the other agents and do not need to be cardinal, an ordinal utility is enough.

Our mechanism starts with each agent building partial solutions which better account for its preferences. Individual partial solutions are then merged by agents, a process in which agents make concessions. A complete solution may eventually be arrived at by a group of agents, thus a coalition is formed. In this model agents seek gradual and reciprocal concessions. Without concessions, solutions will not be reached and the agents will not reach their objectives. The concessions are made gradually. Step by step, each agent is asked to make a compromise that will set its preferences closer to the others'. We re-emphasize here that making concessions is a rational behavior (as suggested in [3]), as it increases the expected utility of the conceding agent. In the example problem domain we study, the criteria agents should make concessions regarding the documents to which they will be applied first. At the first stage of their processing, these agents draw a list of documents they prefer to analyze and give them an order of priority. At the coalition formation stage, agents should modify their preferences regarding either the documents to be analyzed or their priorities. Prioritizing is necessary since applying all the criteria to a document is infeasible, as there are thousands of criteria which could be applied.

### 3.2 Definitions

Before presenting the details of the coalition formation protocol and agent behaviors, we introduce some definitions. We initially define the concepts on which the coalitions and the solutions agents search for are based. We proceed with defining various types of relationships between tasks.

**Definition 1: Coalition**. A set of agents that joined together to perform a task or a combination of tasks.

**Definition 2: Coalition structure**. A set of coalitions that, together, can address all of the tasks and task combinations to be performed at a given moment. If a coalition structure is approved by all agents, it is considered a solution. A solution guarantees that all of the tasks are executed by the agents. The set of tasks and task combinations that the coalitions in the solution will perform is called the *support* of the coalition structure.

**Definition 3: Group of coalition structures**. A set of coalition structures that each provides a specific agent with the same utility. For brevity, it will be referred to as a group of structures or simply a group.

**Definition 4: Context**. A set of parameters which must be stable during a negotiation step.

A context is particularly important in the application domain we address, because the locations of documents change rapidly. Example context parameters are date and time.

**Definition 5: Utility function**. Measures the satisfaction of the agent with, or its surplus from, its collective and individual actions. It is used to represent the preferences of the agents. It may be ordinal or cardinal. In our case, measuring the utility of a structure of coalitions means comparing it with a reference state. The reference state will be the same one throughout the negotiation.

**Definition 6: Reference state**. In order for the agents to know whether they should accept a coalition structure as a solution, they need to be able to compare it with their minimal guaranteed gain during the negotiation. This minimum is the reference state. If there are already formed coalitions, the reference is the current coalitional state.

To guarantee a solution after a negotiation, the reference state needs to be feasible and identical across all of the criteria agents. Otherwise, solutions arrived at by different agents might be inconsistent amongst themselves.

**Definition 7: Acceptable coalition structure**. A coalition structure is acceptable for an agent if it is preferred over, or equivalent, to the current reference state.

**Definition 8: Signature** of a coalition structure $E_i$, denoted $Sig(E_i)$, defines the set of criteria agents that have approved this coalition structure.

**Definition 9: Knowledge** $K(C_i)$ of an agent $C_i$ is the set of coalition structures, and the corresponding signatures,

known to $C_i$.

**Definition 10: Unacceptable** coalition structures: a set of coalition structures $Out(C_i)$ for which agent $C_i$, notifies the other agents of it being unacceptable for it. An unacceptable coalition structure cannot be proposed as a solution by any agent.

The search for the support of a solution considers relationships between tasks. Below, we define some types of these relationships and provide examples. First, we introduce the relationships between tasks by considering the agents which will perform them; then, we consider relationships based on the structure of the tasks.

**Definition 11: Total covering.** Tasks $T_i$ and $T_j$ are in a total covering relationship if the agents selected to carry out $T_i$ can also carry out $T_j$ and inversely, and the tasks do not conflict (resource conflicts, etc.).

In the case of total covering, tasks $T_i$ and $T_j$ could be carried out either sequentially or in parallel, since possible conflicts are avoided.

**Definition 12: Inclusive covering.** Tasks $T_i$ and $T_j$ are in inclusive covering if a subset of the agents selected to carry out one of the two tasks can carry out the other.

**Definition 13: Partial covering.** Tasks $T_i$ and $T_j$ are in partial covering if a subset of the agents selected to carry out $T_i$ can also carry out *a part* of $T_j$ and inversely.

**Definition 14: Total complementary tasks.** Tasks $T_i$ and $T_j$ are totally complementary if each of them can reuse the results of the other task.

As an example, consider two documents $D_i$ and $D_j$ indexed by the same site. Based on preliminary experiments we performed, we know that, (a) frequently, documents found at the same site are similar to one another (in the linguistic sense); (b) because of their similarity, the results obtained when analyzing $D_i$ subject to certain criteria, may be applicable to $D_j$ too. For instance, computing the frequencies of one or several words in a document could be a shared result. This introduces a complementary relationship.

**Definition 15: Dependent tasks.** Tasks $T_i$ and $T_j$ are dependent if only one of the two tasks needs the results of the other task to perform its execution.

The dependence relationship between tasks allows grouping of agents based on their shared interests. For example, in order to analyze a document $D_i$ at a given site, several criteria are necessary, among which some may be relevant to the analysis of another document $D_j$ at the same site. Such criteria sharing may promote the formation of a coalition for performing the two analysis tasks of $D_i$ and $D_j$.

**Definition 16: Competitive tasks.** Tasks $T_i$ and $T_j$ are competitive if $T_i$ and $T_j$ compute the same indicators using different methods. $T_i$ and $T_j$ should not be part of the support of a same solution. Recognizing competitive tasks makes it possible to prune the solutions search space.

## 3.3 The coalition formation method

The aim of this method is to solve the agent coalition formation problem without having to aggregate the preferences of the agents, and to allow a dynamic and fast reorganization of these coalitions according to changes in the problem domain. The method we propose is based on two concurrent behaviors of the agents: task analysis, and negotiation. Task analysis consists of grouping of tasks into combinations, based on relationships among the tasks, where each combination is to be performed by a coalition. To simplify and optimize the search among those combinations of tasks, a binary tree is constructed, and combinations are placed in its nodes. Additionally, from each combination of tasks competitive combinations (defined below) can be isolated. The advantage of the suggested method is in directing the search of the support of the solutions towards preferred tasks, thus reducing search complexity.

**Definition 17: Competitive combination of tasks.** Two combinations of tasks $CT_1$ and $CT_2$ are competitive for an agent $a_i$ if its utility from $CT_1$, $U(a_i, CT_1)$, is equal to its utility from $CT_2$, $U(a_i, CT_2)$, and $CT_1$ and $CT_2$ do not belong to the support of the same solution.

**Definition 18: Tree of task combinations** is constructed based on combinations' preference. The preferred combination of the agent that constructs the tree is placed in the root. The rest of the nodes are populated in a descending preference order. Two branches emanate from each node $N$: a positive branch labeled with (+), indicating that $N$ is in the support of the solution; a negative branch labeled with (-), which indicates that $N$ is out of the support of the solution.

The supports of the coalition structures in this tree are read starting from the root and by retaining only the nodes when branches emanating from them are labeled with a positive sign. For instance, in Figure 3 the support $S_3$ is only formed of the combinations $<\{T_1, T_2, T_3\}, \{T_5, T_6\}>$ since on the path starting from the root to the leaf node $S_3$, the node $\{T_4, T_5\}$ is negatively labeled on the branch of this path. Hence, this node does not belong to the support $S_3$.

The tree of combinations is incrementally constructed. To develop this tree an agent selects the combinations that it organized first in partitions. We define a partition of task combinations as a set of combinations concerned with one task and containing all the combinations where this task appears. Each combination of tasks belongs to only one partition (cf. Figure 2).

For instance, consider four agents $C_1,...,C_4$, eight tasks $T_1,...,T_8$ and seven possible coalition structures $E_0,...,E_6$, where $E_0$ is the initial state. Of these structures, only three are Pareto optima ($E_1$, $E_3$ and $E_6$). Let $U_i(E)$ be the utility of agent $C_i$ for the coalition structure $E$. We represent the coalition structures in Figure 1 according to the utility that they bring to each agent. To build its coalition structures, agent $C_1$ computes some task combinations. This is done with respect to its preferences. For instance, $C_1$ computes the following preferred combinations: *{T$_1$, T$_2$, T$_3$}, {T$_1$, T$_2$}, {T$_4$, T$_5$}, {T$_5$, T$_6$}, {T$_3$, T$_4$, T$_5$}, {T$_6$, T$_7$}* and *{T$_6$, T$_7$, T$_8$}*.

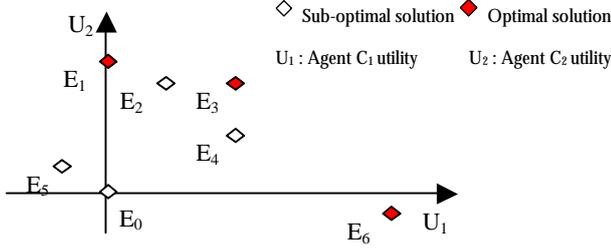

**Figure 1.** Possible coalition structures in a utility space

In this example, these combinations are used to build the coalition structure $E_1$ as explained hereafter. To simplify the construction of, and the search in, the tree of the supports for the solutions, these combinations are first classified into several partitions ($P_1$, $P_2$ and $P_3$ in Figure 2) and ordered in decreasing preference order. The preference degree of each partition is computed based on the preference degrees of the combinations it includes. The first partition $p_1$ has the highest preference degree and is thus placed at the top. In building the tree, the preferred combination of agent $C_1$, *{T$_1$, T$_2$, T$_3$}*, is attached to the root of its tree (cf. Figure 3). Then, the agent creates two branches and searches for the next combination to be considered. In the positive branch, the combination must be selected from a partition other than $P_1$, because combinations in $p_1$ share at least the task $T_1$, and a support of a coalition structure should have only one instance of each task. The agent selects the combination *{T$_4$, T$_5$}* from $P_2$. As for the negative branch, the agent selects the combination *{T$_1$, T$_2$}* from $P_1$. The tree is further built in the same manner. For instance, the following positive combination is *{T$_6$, T$_7$}* form $P_3$.

Note that singleton combinations (containing one task) are not immediately integrated into the tree. To avoid building unnecessarily deeper trees, singletons are directly added to the support once formed. Searching the tree and selecting the nodes from which positively labeled branches emanate gives us already several possible supports, for instance, $S_3$. $S_3$ is formed of two combinations *<{T$_1$, T$_2$, T$_3$}, {T$_5$, T$_6$}>*. If among the singleton combinations, there are some that the agent would like to perform, they are added to the support. E.g., if $C_1$ is interested in tasks *{T$_4$}* and *{T$_7$}*, $S_3$ becomes *<{T$_1$, T$_2$, T$_3$}, {T$_5$, T$_6$}, {T$_4$}, {T$_7$}>*.

Once the agent has identified the supports in the tree of task combinations, and in particular its most preferred coalition structures, it is ready to start the negotiation phase. In particular, it will send coalition structures it has found as part of coalition formation proposals to peer agents. Each agent computes its own tree. Each tree may provide several supports of different coalition structures, and such supports may vary across trees. This in turn allows several different solutions to the coalition formation problem, as each task or combination of tasks in each support can be allocated to different coalitions of agents.

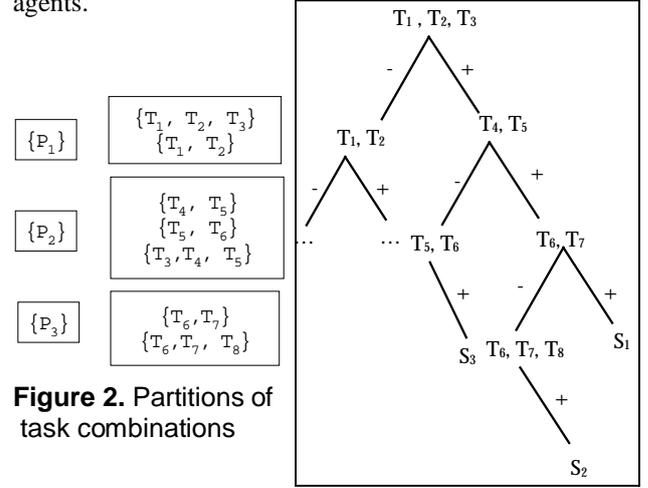

**Figure 2.** Partitions of task combinations

**Figure 3.** Supports tree

**3.3.1 Initiation of the negotiation**. The negotiation process we propose is based on three phases: initialization of the negotiation and transfer of tasks, the core negotiation, and transmission of the solution. There is no pre-established order among the agents. Agents can initiate negotiation and participate at any time. Negotiation may be initiated when new documents to be analyzed are sent by the search engines or by a peer criterion agent. Each agent asks the other agents to send it their tasks, prioritized. Upon such a request, each agent computes and sends the vector of its conditional choices. We assume that agents are cooperative in this respect. For instance, in our application domain, tasks are documents to be analyzed. An agent $C_i$ may send a vector, e.g., $Want(C_i)=(t_1:D_1 \vee D_2 \vee D_3, t_2:D_4$ if $C_j \ D_4, :D_5)$. This vector indicates that $C_i$ wants to analyze documents $D_1$, $D_2$ or $D_3$ first (at time $t_1$). It wants to analyze document $D_4$ second (at time $t_2$), if the necessary resources are provided by agent $C_j$. It has no demands for document $D_5$.

After task vectors with preferences are transmitted, each agent holds the set of tasks and can compute coalitions to be proposed or agreed to. The initiator agent computes the preferred combinations of tasks. It constructs a tree of supports as described above. Each support in this tree will provide a set of coalition structures. Indeed, for each task or combination of tasks in this support, the initiator agent finds its preferred agents which will perform them. The agent then gathers these coalition structures in groups in order to initiate the negotiation.

**3.3.2 The negotiation model.** Once the initiator agent computes the coalition structures, it chooses an agent to which these coalition structures will be sent (as proposals for coalition formation). This choice is based on the agent's strategies. The agent also declares its unacceptable coalition structures. The initiator initially sends its most preferred coalition structures, signed in $Sig(E_i)$; it may iteratively send, in a decreasing order of preference, its other coalition structures. This may be continued until there are no more coalition structures at least equivalent to the current state. However, before sending a less preferred coalition structure, the agent may wait until it receives a message from another agent either about the former coalition structure it has proposed or about a new coalition structure proposed by that agent. Each of the other agents also computes its preferred coalition structures. It then computes the coalition structures that it will choose in second position and so on.

When an agent $R$ receives a group of coalition structures from a sender agent $S$, it sorts those coalition structures in order of preference into homogeneous new groups. In each of these groups, all coalition structures are equivalent in terms of utility of $S$. $R$ updates its knowledge on the coalition structures in $K(R)$. When the utility of a new group is equivalent to the utility of a group which is already known to $R$, these groups are not merged by $R$, since for $S$ their utility is not the same. $R$ only sorts coalition structures that are at least equivalent to the reference state and the others are not considered. If there is at least one coalition structure which is preferable or equivalent to $R$'s best choice, it forwards this coalition structure $CS$ to the next agent that it wishes to include in the negotiation. $R$ signs $CS$ and adds it to $Sig(CS)$. This information indicates that $CS$ has been approved by $R$ and by all the preceding senders as well. When $R$ finds in the set it receives unacceptable coalition structures, it has to declare them unacceptable to the other agents in $Out(R)$. $Out(R)$ also contains the coalition structures that $R$ itself, locally, identified as unacceptable.

Information on unacceptable structures is useful because it prevents the need of other agents computing coalition structures that will be systematically refused. However, its transmission is expected only when agents trust each other. With no trust, agents would not transmit this information, to prevent the others from using the information in their strategies. For instance, such information could enable agents to know which coalition structures an agent may accept, thus gaining an advantage in negotiation. When a combination of tasks in a support of a coalition structure is unacceptable for an agent, it should also indicate that. This combination of tasks is then deleted from the partitions of tasks to prevent it from being considered as part of a support of another coalition structure. The tree is also revised in order to erase the supports containing this combination of tasks.

In case that a coalition structure $CS$ is acceptable for agent $A$, but its utility is inferior to its top choice, $A$ may nevertheless decide to forward $CS$ to the next agent $B$. Consequently the number of agents having approved $CS$ grows and $CS$ is thus reinforced. $A$ may also decide to temporarily block $CS$ but indicates in $K(A)$ that it has received it, if it considers that there is still enough negotiating time to reach a consensus. In this case the agent sends to $B$ the coalition structures it prefers.

A possible end point of the negotiation occurs when an agent $C$ receives a group of coalition structures approved by all other agents. $C$ sorts the coalition structures into groups. If at least one of the coalition structures is better than, or equivalent to, $C$'s reference state, and if negotiation time is about to expire, it can consider the suggested structure as its best group. All the coalition structures of this group are Pareto optima, so $C$ can arbitrarily choose one of them as a solution for the negotiation. In case that negotiation time has not expired and provided that $C$ has some undeclared groups of coalition structures, it can continue the negotiation.

Once the last agent has identified a Pareto optimal solution which is approved by all, it sends this coalition structure to the other agents, which accept it as the solution for the negotiation, as they have already confirmed it in $Sig(E_i)$ and are thus committed to it.

**3.3.3 Negotiation example.** We illustrate the negotiation model using the previous example. For simplicity, we limit our discussion to two agents, $C_1$ and $C_2$. Agent $C_1$ initiates the negotiation and it builds its tree of combinations of tasks. Then $C_1$ generates a set of coalition structures it considers acceptable. It then sorts them into equivalent groups of coalition structures: $G_1(E_6)$ ; $G_2(E_4;E_3)$ ; $G_3(E_2)$ ; $G_4(E_0;E_1)$. In each group, coalition structures have the same preference (cf. Figure 4). $E_5$ is not sorted as the reference state ($E_0$) is better.

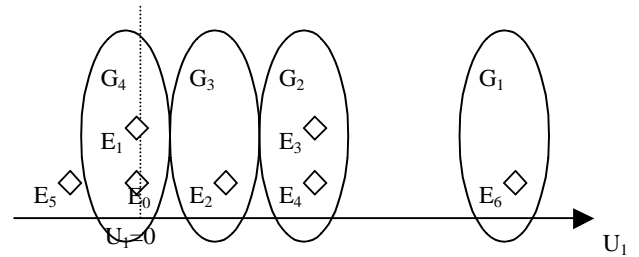

**Figure 4.** Group of coalition structures of agent $C_1$

Groups $G_1$, $G_2$, $G_3$ and $G_4$ are acceptable to agent $C_1$ as they correspond to a state which is as satisfactory as the initial reference state, or better. In the same way, agent $C_2$ also searches for its preferred acceptable coalition structures and sorts them into equivalent groups of coalition structures (Figure 5): $G'_1(E_1)$; $G'_2(E_2;E_3)$; $G'_3(E_4)$; $G'_4(E_5)$; $G'_5(E_0)$. $E_6$ is not sorted as the reference state $E_0$ is better.

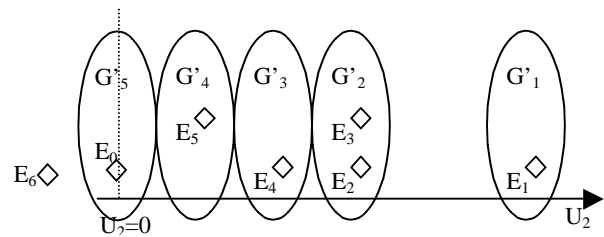

**Figure 5.** Group of coalition structures of agent $C_2$

Groups $G'_1, G'_2, G'_3, G'_4$ and $G'_5$ are acceptable to agent $C_2$ as they correspond to a state which is as satisfactory as the initial reference state, or better. Thus $C_1$ starts by sending its first preferred group $G_1$. Agent $C_2$ starts by receiving $G_1$ and evaluates it (cf. Figure 6). The unique coalition structure $E_6$ in $G_1$ is unacceptable for $C_2$ because it leads to a state less satisfactory than the initial state. Agent $C_2$ does not send this coalition structure. It notifies $C_1$ that it rejects $G_1(E_6)$. If one or more coalitions in the coalition structure $E_6$ is appropriate for $C_2$, it may indicate it to $C_1$.

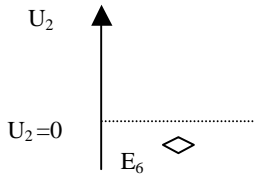

**Figure 6.** First group received by agent $C_2$

Agent $C_1$ must then wait for a new proposal from agent $C_2$. Thus $C_2$ sends its preferred set, i.e. $G'_1(E_1)$. This group contains only coalition structure $E_1$ which is acceptable to agent $C_1$ but which corresponds at the same time to its least preferred choice. For $E_1$ the utility of $C_2$ is the same as that of its reference state $E_0$ (cf. Figure 5). Agent $C_1$ does not reject this coalition structure but it still has a possibility to propose a new group to agent $C_2$. $C_1$ thus sends its second preferred group $G_2(E_3; E_4)$.

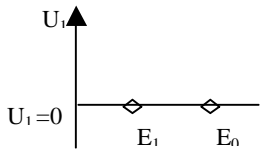

**Figure 7.** First group received by agent $C_1$

Agent $C_2$ receives $G_2$ and separates its two coalition structures into two groups. $E_3$ of $G_2$ now belongs to its group $G'_2$ which corresponds to its second preferred group. As for coalition structure $E_4$, it belongs to group $G'_3$ − its third choice. $E_3$ is acceptable. As all the other agents have already participated in the negotiation, agent $C_2$ cannot send it to others. Coalition structure $E_3$ of $G'_2$ can thus be a solution. Agent $C_2$ has no other Pareto optimal choices left. Either it sends $E_3$ to agent $C_1$ in order to indicate to it the final result of the negotiation, or it waits until $C_1$ gives way on coalition structure $E_1$ which is also Pareto optimal considering the fact that agent $C_2$ has already refused one of the coalition structures that agent $C_1$ proposed to it. The negotiation between the two agents inevitably finishes on one of the two Pareto optima and before expiration of the pre-set negotiation time. Note that classical game theoretic analysis might bring the two agents to an equilibrium where no coalition is formed (and both lose). Yet, as shown in [3], and reinforced in our work, it is in the best interest of agents, even if they are self-interested, to compromise. Implementing a compromise strategy, one of these agents will normally yield to allow for a solution to be arrived at.

## 4 Experimental evaluation

To evaluate it, the model suggested in this study was implemented and experimented with. The code was written in Java and run on a Windows 98 host, 1.4 GHZ processor and 256 MB RAM. We initially performed experiments on the basis of some dependence relationships. In order to evaluate the protocol, we have analyzed its performance by observing several parameters: the number of exchanged messages, the size of these messages (the number of coalition structures they contain), the number of coalition structures that have been evaluated and the negotiation runtime. The results obtained are summarised in the following figures. It should be noted that each point in the graphs is the average of 10 tests carried out under the same conditions.

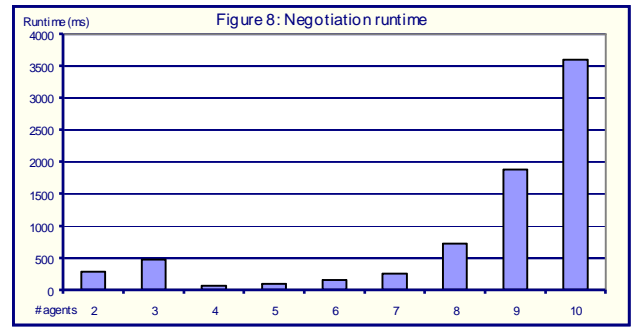

Figure 8 shows negotiation runtimes obtained in milliseconds with the number of agents varying between 2 and 10. The negotiation runtime increases in this figure. This is due to the higher number of proposals that agents would compute and exchange. However the search time remains acceptable even when the number of agents grows. Figure 9 shows the number of coalition structures sent by the agents. For instance in our experiments, with 4 agents and 4 documents, 18 coalition structures were transmitted, and for 10 agents with 10 documents, only 5058 coalition structures have been sent, compared to 45927 coalition structures expected in a case where dependencies are not considered (the latter was computed offline). The number of messages sent varies considerably according to the incompatibility of the preferences of the agents.

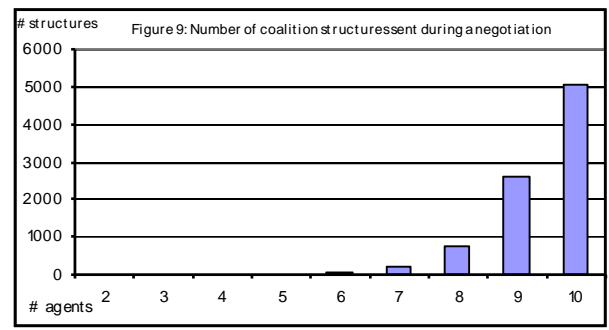

Figure 10 shows the number of evaluated coalition structures and allows measuring the effectiveness of the search when handling of dependence relationships is employed. Had agents not considered dependencies, for instance in the case of 4 agents and 4 tasks, the number of coalition structures they should have evaluated would have been 6561. As the graph shows, the actual number is by far smaller. This holds for larger numbers of agents too: 10 agents have only examined 8983 coalition structures, compared to 45927 possible ones when dependencies are not handled.

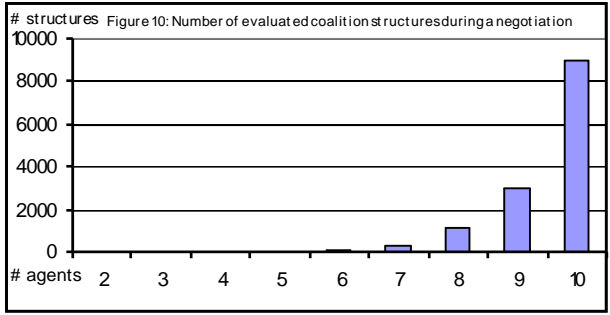

To evaluate the gains from the use of task dependencies as part of the coalition formation process, we performed two experiments where in each the same tasks were provided to the agents, however in one experiment they used task dependencies and in the other they did not. Comparative results between these two experiments are presented in Figure 11. As seen there, the gains of coalition formation with the handling of task dependencies are, on average, about 13% higher than the gains without it. This gain in computation results from task combinations that are facilitated by task dependencies. By using task combinations, the agents reduce the number of coalitions they need to consider.

For instance, without the use of the combinations, an agent that has 4 tasks must examine the 4 corresponding coalitions. But with task combinations, for the same 4 tasks, the number of coalitions may decrease. E.g., if the tasks are gathered in two task combinations, the agent will have to examine only 2 corresponding coalitions.

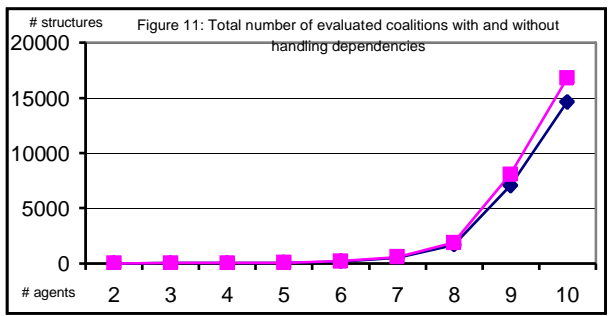

Note that although the results presented here refer to 10 agents, we have performed additional experiments in larger systems, exceeding 50 agents. The subset presented is merely aimed at providing the essence of the results.

## 5 Conclusion

Classical solutions to the coalition formation problem assume cardinal ordering of task values. Additionally, it is commonly assumed that agents behavior rationally, and should thus not compromise their utility. Further, existing solutions hardly analyze complex task dependencies. Recently, it was shown [3] that compromise strategies in coalition formation may dominate other strategies. In this study we leverage on this new result. We advocate that, for some problem domains, compromise is necessary not only to facilitate the formation of coalitions; it is also necessary to reduce the complexity of coalition formation. We further claim that task dependencies can by utilized to prune the coalition formation search space. As we show in our experiments, the use of the compromise-based coalition formation model we present provides significant savings in the computation and communication complexity of coalition formation. Our results also show that when information on task dependencies is used, the complexity of coalition formation is further reduced.

This study should be extended to address several issues. Firstly, the sizes of the systems examined were relatively small. Since in our experiments we observed a steep growth in computational costs, scale-up of the proposed mechanism may face difficulties, in spite of it performing much better than a naïve approach. Secondly, the mechanism should be check in other domains. Although it was specifically designed, and proved successful, for tracking racist documents, it is desirable to prove it applicable to other domains. In future work we intend to pursue these directions. The current results are already very promising, as were able to utilize a MAS technique – coalition formation – for solving an important real-world problem.

## References

[1] Cardie, C. *Empirical Methods in Information Extraction, AI Magazine,* 18(4):65-80, 1997.
[2] Kraus, S., Shehory, O., Tasse, G. *Coalition Formation with Uncertain Heterogeneous Information*, AAMAS, 1-8, 2003.
[3] Kraus, S., Shehory, O., Tasse, G., *The Advantages of Compromising in Coalition Formation with Incomplete Information,* AAMAS, 588-595 2004.
[4] Letsche, T. Berry, M.W. *Large-scale Information Retrieval with Latent Semantic Indexing. Information Sciences*, 100:105–137, 1997.
[5] Sandholm T.W., Lesser V.R. *Coalitions among Computationally Bounded Agents*, AI, 99-137, 1997.
[6] Sandholm T.W., Larson K., Andersson M., Shehory O., Tohmé F. *Coalition Structure Generation with Worst Case Guarantees*, Artificial Intelligence, 111, 209-238, 1999.
[7] Shehory O, Kraus S. *Methods for Task Allocation via Agent Coalition Formation*, AI, 165-200, 1998.